\documentclass{book}
\usepackage{makeidx,elsevierg}
 \usepackage{graphics}
 \usepackage{graphicx}

\usepackage{amssymb}
\usepackage{latexsym}
\newcommand{\uline}{\vrule height.06ex depth.02ex width.6em}
\newcommand{\mvee}{\vee\kern-.69em\uline}

\makeindex
\collection
\begin{document}


\paper{Is Quantum Logic a Logic?}{Mladen Pavi\v ci\'c and Norman D.~Megill}

\section{Introduction}
\label{sec:intro}
Thirty seven years ago, Richard Greechie and Stanley Gudder
wrote a paper entitled {\em Is a Quantum Logic a
Logic?}~\cite{greechie-gudder71} in which they strengthen
a previous negative result of Josef Jauch and Constantin
Piron.~\cite{jauch-pir70}

``Jauch and Piron have considered a possibility that a quantum
propositional system is an infinite valued logic\ldots
and shown that standard propositional systems (that is,
ones that are isomorphic to the lattice of all closed
subspaces of a Hilbert space) are not conditional and
thus cannot be logic in the usual sense.''~\cite{greechie-gudder71}
A {\em conditional} lattice is defined as follows.
We define a valuation $v[a]$ as a mapping from an element
$a$ of the lattice to the interval $[0,1]$.
We say that two elements $a,b$ are conditional if there
exists a unique $c$ such that $v[c]=min\{1,1-v[a]+v[b]\}$.
We call $c$ the {\em conditional} of $a$ and $b$ and
write $c=a\to b$. We say that the lattice is conditional
if every pair $a,b$ is conditional. Greechie and Gudder then
proved that a lattice is conditional if and only if it
contains only two elements 0 and 1.\footnote{We define
0 and 1 in a lattice in Section \ref{sec:latt}.}
This implies that [0,1] reduces to $\{0,1\}$ and that the
lattice reduces to a two-valued Boolean algebra. In effect,
this result shows that one cannot apply the same kind
of valuation to both quantum and classical logics.

It became obvious that if we wanted to arrive
at a proper quantum logic, we should take an axiomatically
defined set of propositions closed under substitutions and
some rules of inference, and apply a model-theoretic approach
to obtain valuations of every axiom and theorem of the logic.
So, a valuation should not be a mapping to [0,1] or $\{0,1\}$
but to the elements of a model. For classical logic, a model
for logic was a complemented distributive lattice, i.e., a
Boolean algebra. For quantum logics the most natural
candidate for a model was the orthomodular lattice, while
the logics themselves were still to be formulated.
Here we come to the question of {\em what logic is}.
We take that logic is about propositions and inferences
between them, so as to form an axiomatic deductive system.
The system always has some algebras as models, and we always
define valuations that map its propositions to elements
of the algebra---we say, the system always has its
semantics---but our definition stops short of taking
semantics to be a part of the system itself.
Our title refers to such a definition of logic, and
we call quantum logic so defined {\em deductive quantum
logic}.\footnote{Note that many authors understand quantum
logic as simply a lattice
\cite{jauch} or a poset \cite{varad,ptak-pulm}.
Quantum logics so defined do not have the aforementioned
valuation and are not deductive quantum logics. Such a
definition stems from an operationalist approach, which
started with the idea that quantum logic might be empirical.
It was argued that propositions might be measured and that
properties such as orthomodularity for quantum systems or
distributivity for classical ones can be experimentally
verified.~\cite{jauch}} Classical logic is deductive in
the same sense.

In the early seventies, a number of results and a number
of predecessors to deductive quantum logics were formulated.
Jauch, Piron, Greechie, and Gudder above assumed the
conditional---from now on we will call it {\em implication}---to
be defined as $a\to_0 b=a'\cup b$ (see Section \ref{sec:latt}
for notation). However, it was already then known that
in an orthomodular lattice,\footnote{The lattice of all closed
subspaces of a Hilbert space is an orthomodular lattice. See
Section~\ref{sec:latt}.} an implication so defined would not satisfy
the condition $a\to b=1\ \Leftrightarrow\ a\le b\,$, which holds
in every Boolean algebra and which was considered plausible
to hold in an orthomodular lattice too. In 1970, the following
implication was found to satisfy this condition:
$a\to_1 b=a'\cup(a\cap b)$ (the so-called {\em Sasaki
hook}\footnote{The Sasaki hook is an orthocomplement to the
{\em Sasaki projection} \cite{sasaki}.}) by Peter Mittelstaedt
\cite{mittelstaedt70} and Peter Finch \cite{finch70}.
The Sasaki hook becomes equal to $a'\cup b$ when 
an orthomodular lattice satisfies the distributive law, i.e., 
when it is a Boolean algebra.
The Sasaki implication first served several
authors simply to reformulate the orthomodular lattice in a
logic-like way and call it ``quantum
logic.''~\cite{finch70,clark73,piziak74} In 1974 Gudrun
Kalmbach proved that in addition to the
Sasaki hook, there are exactly four other ``quantum implications''
that satisfy the above plausible condition and that all
reduce to $a'\cup b$ in a Boolean algebra.

In the very same year, four genuine (i.e.\ propositional)
deductive quantum logics---using three different implications
and none at all, respectively---were formulated
by Gudrun Kalmbach \cite{kalmb74} (a standard propositional
logic based on the {\em Kalmbach implication}\footnote{
Kalmbach implication is defined as $a\to_3b=(a' \cap
b)\cup(a' \cap b')\cup(a\cap(a' \cup b))$.}),
Hermann Dishkant \cite{dishk} (a first-order predicate logic based on
the {\em Dishkant implication}\footnote{Dishkant implication 
is defined as $a\to_2b=b'\to_1a'$.}), Peter Mittelstaedt
\cite{mittelstaedt74} (a dialog logic based on the Sasaki hook), and
Robert Goldblatt \cite{gold74} (a binary logic with no
implication---the binary inference `$\vdash$' represented
the lattice `$\le$'). Several other quantum logics were later
formulated by Maria Luisa Dalla Chiara \cite{dalla-c-77}
(first-order quantum logic), Jay Zeman \cite{zeman78} 
({\em normal logic}), Hirokazu Nishimura \cite{nishimura80} 
(Gentzen sequent logic), George Georgacarakos \cite{gaca80} 
({\em orthomodular logics} based on {\em 
relevance},\footnote{Relevance implication is defined as 
$a\to_5b=(a\cap b)\cup(a'\cap b)\cup(a' \cap b')$.}  
Sasaki, and Dishkant implications), Michael Dunn \cite{dunn}
(predicate binary logic), {Ernst-{W}alter} Stachow
\cite{stachow-tableaux} ({\em  tableaux calculus},
a Gentzen-like {\em calculus of sequents}, and a
Brouwer-like logic), Gary Hardegree \cite{harde81c}
({\em orthomodular calculus}), John Bell \cite{bell-86}
(quantum ``attribute'' logic), Mladen Pavi\v ci\'c 
\cite{pav87} (binary quantum logics with {\em merged 
implications}\footnote{\label{footn:merged}Under {\em merged 
implications} all six implications are meant; $a\to_ib$, 
$i=0,1,2,3,5$ are defined above; $a\to_4 b=b'\to_3a'$ is 
called {\em non-tollens implication}. In these logics of  
Pavi\v ci\'c, axioms of identical form hold for each of the 
implications yielding five quantum logics and one classical (for
$i=0$).}), 
Mladen Pavi\v ci\'c \cite{pav89} (unary quantum logic with 
{\em merged 
implications}),$\!$\footnote{\label{footn:merged-unary}Again, 
axioms of identical form hold for all implications.} Mladen 
Pavi\v ci\'c and 
Norman Megill \cite{mpcommp99} (unary quantum logics with {\em 
merged equivalences}\footnote{\label{footn:merged-equiv}Merged 
equivalences, 
$a\equiv_i b$, $i=0,\ldots,5$, are explicit expressions 
(by means of $\cup,\cap,'$) of $(a\to_ib)\cap(b\to_ja)$, 
$i=0,\ldots,5$, $j=0,\ldots,5$, in any orthomodular lattice
as given by Table 1 of Ref.~\cite{mpcommp99}. In these  
logics, axioms of identical form hold for all equivalences.}), etc.
Logics with the $v(a)=1$ lattice valuation corresponding 
to $\vdash a$ we call {\em unary} logics and logics with 
the $v(a)\le v(b)$ lattice valuation corresponding to
$a\vdash b$ we call {\em binary} logics.

Still, the parallels with classical logic were a major
concern of the researchers at the time. ``I would argue that
a `logic' without an implication \dots  is radically incomplete,
and indeed, hardly qualifies
as a theory of deduction'' (Jay Zeman, 1978).~\cite{zeman78}
So, an extensive search was undertaken in the seventies and
eighties to single out a ``proper quantum implication'' from
the five possible ones on purely logical grounds,\footnote{An
excellent contemporary review of the state of the art was written
in 1979 by Gary Hardegree \cite{harde79}.} but none of the
attempts proved successful.

In 1987 Mladen Pavi\v ci\'c \cite{pav87,pav89} proved that
there is no ``proper quantum implication'' since any one
of the conditions
$a\to_i b=1\ \Leftrightarrow\ a\le b$, 
$i=1,\ldots,5$\footnote{$a\to_ib$, $i=1,\ldots,5$ are defined above. 
See footnotes Nos.~\protect{\ref{footn:merged}} and 
\protect{\ref{footn:merged-unary}}.} \ {\em is} the very 
orthomodularity which, when satisfied by an 
orthocomplemented lattice (the so-called {\em ortholattice}), 
makes it orthomodular. In terms of a logic, the corresponding
logical rules of inference turn
any {\em orthologic} or {\em minimal quantum logic} into a 
quantum logic. He also proved that when the condition
$a\to_0b=1\ \Leftrightarrow\ a\le b$ is
satisfied by an an ortholattice, the lattice becomes a 
complemented distributive one, that is, a Boolean
algebra.\footnote{In any Boolean algebra all six implications
merge.} A corresponding logical rule of inference
turns any orthologic into a classical logic. 

This finding was soon complemented by a proof given by
Jacek Malinowski in 1990 that ``no logic determined by
any class of orthomodular lattices admits the deduction
theorem,''~\cite{malinowski} where the {\em deduction theorem}
says that if we can derive $b$ from $S\bigcup\,\{a\}$ then
we can derive $a\!\to\!b$ from $S$.\footnote{\label{foot:d-t}It
should be stressed here that the deduction theorem is not
essential for classical logic either. It was first proved
by Jaques Herbrand in 1930.~\cite{herbrand30} All classical
logic systems before 1930, e.g., the ones by Whitehead and Russell,
Hilbert, Ackermann, Post, Skolem, \L ukasiewicz, Tarski, etc.,
were formulated without it.}  He also proved that no extension
of quantum logic, i.e., no logic between the quantum and the
classical one, satisfies the deduction theorem.~\cite{mortensen}
The conclusion was: ``Since orthomodular logic is algebraically
well behaved, this perhaps shows that implication is not such
a desirable operation to have.''~\cite{mortensen}

The conjecture was confirmed by Mladen Pavi\v ci\'c in
1993~\cite{pav93}. The above orthomodularity condition does
not require implications.
One can also have it with an essentially
weaker equivalence operation:
$a\equiv b=1\ \Leftrightarrow\ a=b$, where
$a\equiv b=(a\cap b)\cup(a'\cap b')$; we say $a$ and $b$ are
{\em equivalent}.~\cite{pav93,mpcommp99} As above, when this 
condition is satisfied by 
an ortholattice it makes it orthomodular.\footnote{The same 
holds for $a\equiv_ib$, $i=1,\ldots,5$ from footnote 
No.~\ref{footn:merged-equiv}, as well.~\cite{mpcommp99}} Moreover in
any orthomodular lattice $a\equiv b=(a\to_ib)\cap(b\to_ia)$,
$i=1,\ldots,5$. The analogous classical condition
$a\equiv_0 b=1\ \Leftrightarrow\ a=b$, where
$a\equiv_0 b=(a'\cup b)\cap(a\cup b\,')$, amounts to
distributivity: when satisfied by an ortholattice, it makes 
it a Boolean algebra.~\cite{p98,mpcommp99}

On the other hand, it turned out that everything in
orthomodular lattices is sixfold defined: binary operations,
unary operation, variables and even unities and zeros.
They all collapse to standard Boolean operations, variables and
0,1 when we add distributivity. For example, as proved by
Norman Megill and Mladen Pavi\v ci\'c \cite{mpqo01}
$0_{1(a,b)}=a\cap(a'\cup b)\cap(a\cup b\,')$,$\ldots$,%
$0_{5(a,b)}=(a\cup b)\cap(a\cup b\,')\cap(a'\cup b)\cap(a'\cup b\,')$;
$a\equiv_3b=(a'\cup b)\cap(a\cup(a'\cap b\,'))$;
etc.~\cite{mpqo02} Moreover, we can express any of such
expressions by means of every appropriate other in a huge although
definite number of equivalence classes.~\cite{mpqo02} For example,
a shortest expression for $\cup$ expressed by means of quantum
implications is
$a\cup b=(a\to_i b)\to_i (((a\to_i b)\to_i (b\to_i a))\to_i a)$,
$i=1,\ldots,5$.~\cite{mpqo01,mpqo02,mpijtp98,mpijtp03}

For such a ``weird'' model, the question emerged as
to whether it is possible to formulate a proper deductive quantum logic
as a general theory of inference and how independent of its
model this logic can be. In other words, can such a logic be more
general than its orthomodular model?

The answer turned out to be affirmative. In 1998
Mladen Pavi{\v c}i{\'c} and Norman Megill showed that the
deductive quantum logic is not only more general but also
very different from their models.~\cite{mphpa98,mpcommp99}
They proved that
\begin{itemize}
\item Deductive quantum logic is not orthomodular.
\item Deductive quantum logic has models that are ortholattices
       that are not orthomodular.
\item Deductive quantum  logic is sound and complete under these models.
\end{itemize}

This shows that quantum logic is not much different from the 
classical one since they also
proved that \cite{mpcommp99}
\begin{itemize}
\item Classical logic is not distributive.\footnote{Don't be alarmed.
This is {\em not} in contradiction with anything in the literature.
The classical logic still stands intact, and the fact that it is
not distributive is just a feature of classical logic
that---due to Boole's heritage---simply has not occurred to anyone
as possible and which therefore has not been discovered before.
See the proof of Theorem \ref{th:soundness-c},
Theorem \ref{th:non-distr-c}, Lemma \ref{L:equality-d}, and the
discussion in Section \ref{sec:discussion}.}
\item Classical logic has models that are ortholattices
      that are not orthomodular and therefore also not distributive.
\item Classical logic is sound and complete under these models.
\end{itemize}

These remarkably similar results reveal that quantum
logic is a logic in the very same way in which classical logic is a
logic. In the present chapter, we present these results in some detail.

The chapter is organized as follows. In Section \ref{sec:latt}, we
define the ortholattice, orthomodular lattice, complemented
distributive lattice (Boolean algebra), weakly orthomodular lattice 
WOML (which is not necessarily orthomodular), weakly distributive 
lattice WDOL (which is not necessarily either distributive or 
orthomodular), and some results that connect the lattices.
In Section \ref{sec:logic}, we define quantum and classical logics.
In Sections \ref{sec:sound-q} and \ref{sec:sound-c}, we prove the soundness
of quantum logic for WOML and of classical logic for WDOL, respectively.
In Sections \ref{sec:compl-ql} and \ref{sec:compl-cl}, we prove the
completeness of the logics for WOML and WDOL, respectively.
In Sections \ref{sec:compl-oml} and \ref{sec:last}, we prove the
completeness of the logics for OML and Boolean algebra, respectively,
and show that the latter proofs of completeness introduce
hidden axioms of orthomodularity and distributivity in the
respective Lindenbaum algebras of the logics.
In Section \ref{sec:discussion}, we discuss the obtained results.

\section{Lattices}
\label{sec:latt}

In this section, we introduce two models for deductive quantum
logic, orthomodular lattice and WOML, and two models for classical
logic, Boolean algebra and WDOL. They are gradually defined as follows.

There are two equivalent ways to define a lattice: as a partially
ordered set (poset)\footnote{Any two elements $a$ and $b$ of the
poset have a least upper bound $a\cup b$---called
{\em join}---and a greatest lower bound $a\cap b$---called
{\em meet}.} \cite{maeda} or as an algebra~\cite[II.3.$\>${\em Lattices
as Abstract Algebras\/}]{birk2nd}. We shall adopt the latter approach.

\begin{definition}\label{def:ourOL}
An {\em ortholattice}, {\rm OL\/}, is an algebra
$\langle{\mathcal{OL}}_0,',\cup,\cap\rangle$
such that the following conditions are satisfied for any
$a,b,c\in \,{\mathcal{OL}}_0$ {\rm \cite{mpqo02}}:
\begin{eqnarray}
&&a\cup b\>=\>b\cup a\label{eq:aub}\\
&&(a\cup b)\cup c\>=\>a\cup (b\cup c)\\
&&a''\>=\>a\label{eq:notnot}\\
&&a\cup (b\cup b\,')\>=\>b\cup b\,'\\
&&a\cup (a\cap b)\>=\>a\\
&&a\cap b\>=\>(a'\cup b\,')'\label{eq:aAb}
\end{eqnarray}
In addition, since $a\cup a'=b\cup b\,'$ for any $a,b\in
\,{\mathcal{OL}}_0$, we define:
\begin{eqnarray}
\qquad 1{\buildrel\rm def\over=}a\cup a',\qquad\qquad
 0{\buildrel\rm def\over =}a\cap a'\label{D:onezero}
\end{eqnarray}
and
\begin{eqnarray}
\qquad a\le b\ \quad{\buildrel\rm def\over\Longleftrightarrow}\quad\ a\cap b=a
\quad\Longleftrightarrow\quad a\cup b=b
\end{eqnarray}
\end{definition}

Connectives  $\to_1$ ({\em quantum implication}, {\em Sasaki hook}),
$\to_0$ ({\em classical implication}), $\equiv $ ({\em quantum
equivalence}), and $\equiv_0$ ({\em classical equivalence})
are defined as follows:

\begin{definition}\label{def:impl-L}
\qquad $a\to_1 b\ \ {\buildrel\rm def\over =}\ \
a'\cup(a\cap b),\quad\qquad a\to_0 b\ \ {\buildrel\rm def\over =}\ \
a'\cup b$.
\end{definition}

\begin{definition}\label{L:id-bi-L}$\!\!\!$\footnote{In every
orthomodular lattice $a\equiv b=(a\to_1 b)\cap(b\to_1 a)$,
but not in every ortholattice.}
\qquad $a\equiv b\ \
{\buildrel\rm def\over =}\ \ (a\cap b)\cup(a'\cap b\,')$.
\end{definition}

\begin{definition}\label{L:id-bi-C}
\quad\qquad $a\equiv_0 b\ \
{\buildrel\rm def\over =}\ \ (a\to_0 b)\cap(b\to_0 a)$.
\end{definition}

Connectives bind from weakest to strongest in the order $\to_1$ ($\to_0$),
$\equiv $ ($\equiv_0$), $\cup$, $\cap$, and $'$.

\begin{definition}\label{def:woml2} {\rm (Pavi\v ci\'c and Megill \cite{mpcommp99})}
An ortholattice that satisfies the following condition:
\begin{eqnarray}
a\equiv  b=1\qquad \Rightarrow \qquad
(a\cup c)\equiv(b\cup c)=1\label{eq:woml2}
\end{eqnarray} 
is called a {\em weakly orthomodular ortholattice}, {\rm WOML}.
\end{definition}

\begin{definition}\label{def:oml2} {\rm (Pavi\v ci\'c \cite{pav93})}
An ortholattice that satisfies the following condition:
\begin{eqnarray}
a\equiv b=1\qquad\Rightarrow\qquad a=b,\label{eq:oml2}
\end{eqnarray}
is called an {\em orthomodular lattice}, {\rm OML}.
\end{definition}

Equivalently:
\begin{definition}\label{def:oml2o} {\rm (Foulis \cite{foulis62},
Kalmbach \cite{kalmb74})}
An ortholattice that satisfies either of the following two conditions:
\begin{eqnarray}
&&a\cup(a'\cap(a\cup b))=a\cup b\label{eq:oml2o1}\\
&&a\,{\mathcal{C}}\,b\quad\&\quad a\,{\mathcal{C}}\,c
\quad\Rightarrow\quad
a\cap(b\cup c)=(a\cap b)\cup(a\cap c)\label{eq:oml2o2}
\end{eqnarray} 
where $a\,{\mathcal{C}}\,b\ \ {\buildrel\rm def\over\Longleftrightarrow}\
\ a=(a\cap b)\cup(a\cap b\,')$
($a$ {\em commutes} with $b$), is called an {\em orthomodular lattice}, {\rm OML}.
\end{definition}

\begin{definition}\label{def:wdol2} {\rm (Pavi\v ci\'c and
Megill \cite{mpcommp99})}
An ortholattice  that satisfies the following:\footnote{This 
condition is known as
{\em commensurability}.~\cite[Definition (2.13), 
p.~32]{mittelstaedt70}
Commensurability is a weaker form
of the commutativity from Definition \ref{def:oml2o}. Actually, a
metaimplication from commensurability to commutativity
is yet another way to express orthomodularity. They coincide in
any {\rm OML}.}
\begin{eqnarray}
(a\equiv b)\cup(a\equiv b')=
(a\cap b)\cup(a\cap b')\cup(a'\cap b)\cup(a'\cap b')=1
\label{eq:wdol2}
\end{eqnarray} 
is called a {\em weakly distributive ortholattice}, {\rm WDOL}.
\end{definition}

\begin{definition}\label{def:ba2} {\rm (Pavi\v ci\'c \cite{p98})} An
ortholattice that satisfies the following condition:
\begin{eqnarray}
a\equiv_0 b=1\qquad\Rightarrow\qquad a=b\label{eq:ba2}
\end{eqnarray}
is called a {\em Boolean algebra}.
\end{definition}

Equivalently:
\begin{definition}\label{def:ba2o} {\rm (Schr{\"o}der \cite{schroeder})} An
ortholattice that satisfies the following condition:
\begin{eqnarray}
a\cap(b\cup c)=(a\cap b)\cup(a\cap c)\label{eq:ba2ao}
\end{eqnarray} 
is called a {\em Boolean algebra}.
\end{definition}

The opposite directions in Eqs.~(\ref{eq:oml2}) and (\ref{eq:ba2})
hold in any OL.

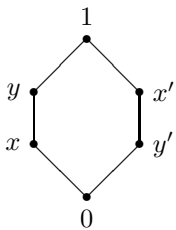
\begin{figure}[htbp]\centering
  \begin{picture}(60,80)(-10,-10)

    \put(20,0){\line(-1,1){20}}
    \put(20,0){\line(1,1){20}}
    \put(0,20){\line(0,1){20}}
    \put(40,20){\line(0,1){20}}
    \put(0,40){\line(1,1){20}}
    \put(40,40){\line(-1,1){20}}

    \put(20,-5){\makebox(0,0)[t]{$0$}}
    \put(-5,20){\makebox(0,0)[r]{$x$}}
    \put(45,20){\makebox(0,0)[l]{$y'$}}
    \put(-5,40){\makebox(0,0)[r]{$y$}}
    \put(45,40){\makebox(0,0)[l]{$x'$}}
    \put(20,65){\makebox(0,0)[b]{$1$}}

    \put(20,0){\circle*{3}}
    \put(0,20){\circle*{3}}
    \put(40,20){\circle*{3}}
    \put(0,40){\circle*{3}}
    \put(40,40){\circle*{3}}
    \put(20,60){\circle*{3}}

  \end{picture}
\caption{Ortholattice O6\label{fig:O6}, also called {\em benzene ring}
and {\em hexagon}.}
\end{figure}

Any finite lattice can be represented by a {\rm Hasse diagram}
that consists of points ({\em vertices}) and lines ({\em edges}).
Each point represents an element of the lattice, and positioning
element $a$ above element $b$ and connecting them with a line
means $a\le b$. For example, in Figure \ref{fig:O6} we have
$0\le x\le y\le 1$. We also see that in this lattice, e.g.,
$x$ does not have a relation with either $x'$ or $y'$.

Definition \ref{df:o6} and Theorems \ref{th:o-O6} and \ref{th:O6}
will turn out to be crucial for the completeness proofs of both
quantum and classical logics in Sections \ref{sec:compl-ql} and
\ref{sec:compl-cl}.

\begin{definition}\label{df:o6} We define {\rm O6} as the lattice
shown in Figure \ref{fig:O6}, with the meaning $0<x<y<1$ and 
$0<y'<x'<1$,
\end{definition}

\begin{theorem}\label{th:o-O6}
An ortholattice is orthomodular if only if it does not include a
subalgebra isomorphic to the lattice {\rm O6}.
\end{theorem}

\begin{proof} Samuel Holland \cite{holl70}. See also
Gudrun Kalmbach \cite[p.~22]{kalmb83}.\end{proof}

\begin{corollary}\label{cor:o6} {\rm O6} violates the distributive law.
\end{corollary}

\begin{proof} Distributivity implies orthomodularity. We can also
easily verify on the diagram:
$y\cap (x\cup x')=y\cap 1=y$, but $(y\cap x)\cup(y\cap x')=x\cup 0=x$.
\end{proof}

\begin{theorem}\label{th:O6} All conditions of {\rm WOML} and {\rm WDOL}
hold in {\rm O6}.
\end{theorem}

\begin{proof} As given by Mladen Pavi{\v c}i{\'c} and Norman
Megill.~\cite{mphpa98,mpcommp99} It boils down to the fact
that {\rm O6} violates none of the conditions given by
Eqs.~(\ref{eq:aub}-\ref{eq:aAb}), (\ref{eq:woml2}), and
(\ref{eq:wdol2}) \end{proof}

\begin{theorem}\label{th:wO6} There exist {\rm WDOL} lattices that are
not orthomodular and therefore not distributive, {\rm WOML} lattices
that are not orthomodular, ortholattices that are neither
{\rm WOML} nor {\rm WDOL}, and there are {\rm WOML} lattices that are not
{\rm WDOL}.
\end{theorem}

\begin{proof} As given by Mladen Pavi{\v c}i{\'c} and Norman
Megill.~\cite{mphpa98,mpcommp99}.\end{proof}

On the one hand, the equations that hold in OML and Boolean algebra
properly include those that hold in WOML and WDOL, since WOML and WDOL
are strictly more general classes of algebras.  But on  the other
hand, there is also a
sense in which the equations of WOML and WDOL can be considered to
properly include those of OML and Boolean algebra, via mappings
that the next theorems describe.

\begin{theorem}\label{th:oml-sim}
The equational theory of {\rm OML}s can be simulated
by a proper subset of the equational theory of {\rm WOML}s.
\end{theorem}

\begin{proof}
The equational theory of OML consists of equality
conditions, Eqs.~(\ref{eq:aub})--(\ref{eq:aAb}) together
with the orthomodularity condition Eq.~(\ref{eq:oml2o1})
(or Eq.~(\ref{eq:oml2}) or Eq.~(\ref{eq:oml2o2})).
We construct a mapping from these conditions to WOML 
conditions as follows. We map each of the OML conditions, 
which is an equation in the form $t=s$ (where $t$ and $s$ are
terms), to the equation $t\equiv s=1$, which holds in
WOML. Any equational proof in OML can then be simulated in
WOML by replacing each axiom reference in the OML proof with
its corresponding WOML mapping.~\cite{pmjlc08}
Such a mapped proof will use only a proper subset of the equations
that hold in WOML: any equation whose right-hand side
does not equal 1, such as $a=a$, will never be used.
\end{proof}

\begin{corollary}\label{cor:w-t}
No set of equations of the form $t\equiv s=1$, where $t$ and $s$
are terms in {\rm OML} and where $t=s$ holds in {\rm OML},
determines an {\rm OML} when added to the conditions for 
ortholattices.
\end{corollary}

\begin{proof}
Theorem \ref{th:oml-sim} shows that all equations of this
form hold in a WOML and none of WOML conditions given by
Eqs.~(\ref{eq:aub}-\ref{eq:aAb},$\>$\ref{eq:woml2}) is violated
by O6. Hence, Theorem \ref{th:o-O6} completes the proof.
\end{proof}

\begin{theorem}\label{th:ba-sim}
The equational theory of Boolean algebras can be simulated
by a proper subset of the equational theory of {\rm WDOL}s.
\end{theorem}

\begin{proof}
The equational theory of Boolean algebras
consists of equality conditions Eqs.~(\ref{eq:aub})--(\ref{eq:aAb})
together with the distributivity condition Eq.~(\ref{eq:ba2ao}).
We construct a mapping from these conditions into WDOL as follows.
We map each of the Boolean algebra conditions, which is an
equation in the form $t=s$ (where $t$ and $s$ are
terms), to the equation $t\equiv_0 s=1$, which holds in WDOL.
Any equational proof in a Boolean algebra can then be simulated
in WDOL by replacing each condition reference in the Boolean
algebra proof with its corresponding WDOL mapping.~\cite{pmjlc08}
Such a mapped proof will use only a proper subset of the equations
that hold in WDOL: any equation whose right-hand side
does not equal 1, such as $a=a$, will never be used.
\end{proof}

\begin{corollary}\label{cor:w-t-d}
No set of equations of the form $t\equiv_0 s=1$, where $t$ and $s$
are terms in any Boolean algebra and where $t=s$ holds in the
algebra, determines a Boolean algebra when added to an ortholattice.
\end{corollary}

\begin{proof}
Theorem \ref{th:ba-sim} shows that all equations of this
form hold in a WDOL and none of WDOL conditions given by
Eqs.~(\ref{eq:aub}-\ref{eq:aAb},$\>$\ref{def:wdol2}) is violated
by O6. Hence, Corollary \ref{cor:o6} completes the proof.
\end{proof}

\section{Logics}
\label{sec:logic}

Logic, $\mathcal{L}$, is a language consisting of
propositions and a set of conditions
and rules imposed on them called axioms and rules of inference.

The propositions we use are well-formed formulas (wffs),
defined as follows.
We denote elementary, or primitive, propositions by
$p_0,p_1,p_2,\dots $, and have the following primitive connectives:
$\neg$ (negation) and $\vee$ (disjunction).
The set of wffs is defined recursively as follows:
\begin{enumerate}
\item[] $p_j$ is a wff for $j=0,1,2,\dots $
\item[] $\neg A$ is a wff if $A$ is a wff.
\item[] $A\vee B$ is a wff if $A$ and $B$ are wffs.
\end{enumerate}

We introduce conjunction with the following definition:

\begin{definition} \label{D:conj}
$A\wedge B\ {\buildrel\rm def\over =}\ \neg (\neg A\vee\neg B)$.
\end{definition}

The statement calculus of our
 metalanguage consists of axioms and rules from the object language
as elementary metapropositions and of compound metapropositions built up
by means of the following metaconnectives: $\sim$ (\it not\rm ),
\& (\it and\rm ), $\mvee$ (\it or\rm ), $\Rightarrow$
(\it if\dots ,\ then\rm ), and $\Leftrightarrow$ (\it iff\rm ), with the
usual \it classical \rm meaning. Our metalanguage
statement calculus is actually the very same classical logic we deal 
with in this chapter, only with the \{0,1\}\ valuation.
We extend the statement calculus of the metalanguage with first-order
predicate calculus---with quantifiers $\forall$ (\it for
all\rm ) and $\exists$ (\it exists\rm )---and informal set theory in the
usual way.

\medskip
The operations of implication are the following ones (classical, Sasaki,
and Kalmbach) \cite{pav87}:

\begin{definition} \label{def:impl-0}
$\qquad A\to_0 B\ \ {\buildrel\rm def\over =} \ \
\neg A\vee B$.
\end{definition}

\begin{definition} \label{def:impl-1}
$\qquad A\to_1 B\ \ {\buildrel\rm def\over =} \ \
\neg A\vee (A\wedge B)$.
\end{definition}

\begin{definition} \label{def:impl-3}
$\qquad A\to_3 B\ \ {\buildrel\rm def\over =} \ \
(\neg A\wedge B)\vee(\neg A\wedge\neg B)\vee(A\wedge(\neg A\vee B))$.
\end{definition}

We also define the {\em equivalence} operations as follows:

\begin{definition}\label{L:equiv}
$\qquad A\equiv B\ \
{\buildrel\rm def\over =}\ \ (A\wedge B)\vee(\neg A\wedge\neg B)$.
\end{definition}

\begin{definition}\label{L:equiv-0}
$\qquad A\equiv_0 B\ \
{\buildrel\rm def\over =}\ \ (A\to_0 B)\wedge(B\to_0 A)$.
\end{definition}

Connectives bind from weakest to strongest in the order $\to$,
$\equiv$, $\vee$, $\wedge$, $\neg$.

\smallskip
Let $\mathcal{F}^\circ$ be the set of all propositions, i.e.,
of all wffs.
Of the above connectives, $\vee$ and $\neg$ are primitive ones.
Wffs containing $\vee$ and $\neg$ within logic $\mathcal{L}$
are used to build an algebra
${\mathcal{F}}=\langle {\mathcal{F}}^\circ,\neg,\vee\rangle$.
In $\mathcal{L}$, a set of axioms and rules of inference are imposed on
${\mathcal{F}}$. {}From a set of axioms by means of rules of inference,
we get other expressions which we call theorems. Axioms themselves
are also theorems.
A special symbol $\vdash$ is used to denote the set of theorems.
Hence $A\in\ \vdash$ iff $A$ is a theorem. The statement
$A\in\ \vdash$ is usually written as $\vdash A$. We read this: ``$A$ is
provable'' since if $A$ is a theorem, then there is a proof for it.
We present the axiom systems of our propositional logics
in schemata form (so that we dispense with the rule of
substitution).

\subsection{Quantum Logic}
\label{subsec:q-logic}

All unary quantum logics we mentioned in the Introduction are
equivalent. Here we present Kalmbach's quantum logic because it is
the system which has been investigated in the greatest detail
in her book \cite{kalmb83} and elsewhere \cite{kalmb74,mphpa98}.
Quantum logic, $\mathcal{QL}$, is defined as a language
consisting of propositions and connectives (operations) as introduced
above, and the following axioms and a rule of inference.
We will use $\vdash_\mathcal{QL}$ to denote provability from
the axioms and rule of $\mathcal{QL}$ and omit the subscript when
it is clear from context (such as in the list of axioms that follow).

\smallskip
\noindent{\bf Axioms}
\begin{eqnarray}
{\rm A1}\qquad &&\vdash A\label{eq:kalmb-A1}\equiv A\\
{\rm A2}\qquad &&\vdash A\equiv B\rightarrow_0(B\equiv C\rightarrow_0
     A\equiv C)\\
{\rm A3}\qquad &&\vdash A\equiv B\rightarrow_0\neg A\equiv \neg B\\
{\rm A4}\qquad &&\vdash A\equiv B\rightarrow_0A\wedge C\equiv B\wedge C\\
{\rm A5}\qquad &&\vdash A\wedge B\equiv B\wedge A\\
{\rm A6}\qquad &&\vdash A\wedge (B\wedge C)\equiv (A\wedge B)\wedge C\\
{\rm A7}\qquad &&\vdash A\wedge (A\vee B)\equiv A\\
{\rm A8}\qquad &&\vdash\neg A\wedge A\equiv(\neg A\wedge A)\wedge B \\
{\rm A9}\qquad &&\vdash A\equiv\neg\neg A\\
{\rm A10}\qquad &&\vdash\neg(A\vee B)\equiv\neg A\wedge\neg B\\
{\rm A11}\qquad &&\vdash A\vee(\neg A\wedge(A\vee B))\equiv A\vee B\label{eq:a11}\\
{\rm A12}\qquad &&\vdash (A\equiv B)\equiv(B\equiv A)\\
{\rm A13}\qquad &&\vdash A\equiv B\rightarrow_0(A\rightarrow_0 B)\\
{\rm A14}\qquad &&\vdash (A\to_0B)\to_3(A\to_3(A\to_3B))\\
{\rm A15}\qquad &&\vdash (A\to_3B)\to_0(A\to_0 B)\label{eq:kalmb-A15}
\end{eqnarray}
{\bf Rule of Inference} ({\em Modus Ponens})
\begin{eqnarray}
\ {\rm R1}\qquad &&\vdash A \quad \& \quad \vdash A \rightarrow_3 B
\quad\Rightarrow \quad \vdash B\label{eq:kalmb-R1}
\end{eqnarray}
In Kalmbach's presentation, the connectives $\vee$, $\wedge$, and $\neg$
are primitive.  In the base set of any model (such as an OML or WOML
model) that belongs to OL, $\cap$ can be defined
in terms of $\cup$ and $'$, as justified by DeMorgan's laws, and
thus the corresponding $\wedge$ can be defined in terms of
$\vee$ and $\neg$ (Definition \ref{D:conj}).  We shall do this for
simplicity.  Regardless of whether
we consider $\wedge$ primitive or defined,
we can drop axioms A1, A11, and A15 because it has been
proved that they are redundant, i.e., can be derived from the
other axioms.~\cite{mphpa98} Note that A11 is what we would
expect to be {\em the} orthomodularity\footnote{Cf.~Definition 
(\ref{def:oml2o}), Eq.~(\ref{eq:oml2o1})}---see 
Eq.~(\ref{eq:a11-1}) and the discussion following the equation.

\begin{definition}\label{D:gamma-ql}
For\/ $\Gamma\subseteq{\mathcal{F}}^\circ$ we say $A$ is derivable from\/
$\Gamma$ and write\/ $\Gamma\vdash_\mathcal{QL} A$ or just\/
$\Gamma\vdash A$ if there is a sequence of
formulas ending with $A$, each of which is either one of the axioms of
$\mathcal{QL}$ or is a member of\/ $\Gamma$ or is obtained from its
precursors with the help of a rule of inference of the logic.
\end{definition}

\subsection{Classical Logic}
\label{subsec:cl-logic}

We make use of the PM classical logical system $\mathcal{CL}$
(Whitehead and Russell's \it Principia Mathematica\/ \rm axiomatization
in Hilbert and Ackermann's presentation \cite{hilb-ack-book} but
in schemata form so that we dispense with their rule of
substitution).  In this system, the connectives $\vee$ and $\neg$
are primitive, and the $\to_0$ connective shown in the axioms is
implicitly understood to be expanded according to its definition.
We will use $\vdash_\mathcal{CL}$ to denote provability from
the axioms and rule of $\mathcal{CL}$, omitting the subscript when
it is clear from context.

\smallskip
\noindent{\bf Axioms}
\begin{eqnarray}
{\rm A1}\qquad &&\vdash A\vee A\to_0 A\label{eq:cl-a1}\\
{\rm A2}\qquad &&\vdash A\to_0 A\vee B\\
{\rm A3}\qquad &&\vdash A\vee B\to_0 B\vee A\\
{\rm A4}\qquad &&\vdash (A\to_0 B)\to_0(C\vee A\to_0 C\vee B)\label{eq:cl-a4}
\end{eqnarray}
{\bf Rule of Inference} ({\em Modus Ponens})
\begin{eqnarray}
{\rm R1}\qquad &&\vdash A \qquad \&\qquad
A\to_0 B\qquad\Rightarrow\qquad\vdash B\label{eq:cl-r1}
\end{eqnarray}

We assume that the only legitimate way of inferring theorems in
$\mathcal{CL}$ is by means of these axioms and the Modus
Ponens rule. We make no assumption about valuations of the 
primitive propositions from which wffs are built, but instead 
are interested in wffs that are valid
in the underlying models.  Soundness and completeness will
show that those theorems that can be inferred from the
axioms and the rule of inference are exactly those that are valid.

We define derivability in ${\mathcal{CL}}$,
$\Gamma\vdash_\mathcal{CL} A$ or just $\Gamma\vdash A$, in the
same way as we do for system ${\mathcal{QL}}$.

\section{The soundness of $\mathcal{QL}$: orthomodularity lost}
\label{sec:sound-q}

In this section we show that the syntax of $\mathcal{QL}$
does not correspond to the syntax of an orthomodular lattice.
We do this by proving the soundness of $\mathcal{QL}$ for WOML.
To prove soundness means to prove that all axioms as well
as the rules of inference (and therefore all theorems) of
$\mathcal{QL}$ hold in its models. Since by Theorem \ref{th:oml-sim}
WOML properly includes OML, proving the soundness of $\mathcal{QL}$
for OML would not tell us anything new, and we can dispense with it.

\begin{definition}\label{exists-c}We call ${\mathcal{M}}=\langle
{\mathcal{L}},h\rangle$ a model
if ${\mathcal{L}}$ is an algebra and
$h:{\mathcal{F}}^\circ\longrightarrow{\mathcal{L}}$, called a valuation,
is a morphism of formulas ${\mathcal{F}}^\circ$
into ${\mathcal{L}}$, preserving the operations $\neg,\vee$
while turning them into $',\cup$.
\end{definition}

Whenever the base set $\mathcal{L}$ of a model belongs to WOML
(or another class of algebras), we say (informally) that
the model belongs to WOML (or the other class).  In particular,
if we say ``for all models in WOML'' or ``for all WOML models,'' we
mean for all base sets in WOML and for all valuations on each
base set.  The term ``model'' may refer either to a specific
pair $\langle{\mathcal{L}},h\rangle$ or to all possible such
pairs with the base set $\mathcal{L}$, depending on context.

\begin{definition}\label{one}We call a formula
$A\in{\mathcal{F}}^\circ$ valid in the model $\mathcal{M}$,
and write $\vDash_\mathcal{M} A$,
if $h(A)=1$ for all valuations $h$ on the model, i.e. for
all $h$ associated with
the base set $\mathcal{L}$ of the model.
We call a formula
$A\in{\mathcal{F}}^\circ$ a consequence of\/
$\Gamma\subseteq{\mathcal{F}}^\circ$
 in the model $\mathcal{M}$
and write $\Gamma\vDash_\mathcal{M} A$
if $h(X)=1$ for all $X$ in\/ $\Gamma$ implies $h(A)=1$,
for all valuations $h$.
\end{definition}

\medskip

For brevity, whenever we do not make it explicit, the notations
$\vDash_\mathcal{M} A$ and $\Gamma\vDash_\mathcal{M} A$ will always
be implicitly quantified over all
models of the appropriate type, in this section for all WOML models
$\mathcal{M}$.  Similarly, when we say ``valid'' without qualification,
we will mean valid in all models of that type.

We now prove the soundness of quantum logic by means of
WOML, i.e., that if $A$ is a theorem in ${\mathcal{QL}}$,
then $A$ is valid in any WOML model.

\begin{theorem}\label{th:soundness}{\rm [Soundness]}
$\qquad\Gamma\vdash A\quad\Rightarrow\quad\Gamma\vDash_\mathcal{M} A$
\end{theorem}

\begin{proof}
We must show that
any axiom A1--A15, given by Eqs.~(\ref{eq:kalmb-A1}--\ref{eq:kalmb-A15}),
is valid in any WOML model $\mathcal{M}$, and that any set of formulas that
are consequences of $\Gamma$
in the model are closed under the rule of inference
R1, Eq.~(\ref{eq:kalmb-R1}).

Let us put $a=h(A)$, $b=h(B)$, \ldots

By Theorem \ref{th:oml-sim}, we can prove that WOML is
equal to OL restricted to all orthomodular
lattice conditions of the form $t\equiv s=1$, where
$t$ and $s$ are terms (polynomials) built from the
ortholattice operations and $t=s$ is an equation that
holds in all OMLs.\end{proof}

Hence, mappings of $\mathcal{QL}$ axioms and its rule of
inference can be easily proved to hold in WOML. Moreover,
mappings of A1,A3,A5--A13,A15 and R1 hold in any ortholattice.
In particular, the
\begin{eqnarray}
{\rm A11\ \ mapping:}\qquad\qquad 
(a\cup(a'\cap(a\cup b)))\equiv (a\cup b)=1
\label{eq:a11-1}
\end{eqnarray}
holds in every ortholattice and A11 itself is redundant, i.e.,
can be be inferred from other axioms. Notice that by
Corollary~\ref{cor:w-t}, $a\equiv b=1$ does not imply $a=b$.
In particular, Eq.~(\ref{eq:a11-1}) does not
imply $(a\cup(a'\cap(a\cup b)))=(a\cup b)$

\section{The soundness of $\mathcal{CL}$: distributivity lost}
\label{sec:sound-c}

In this section we show that the syntax of $\mathcal{CL}$
does not correspond to the syntax of a Boolean algebra.
In a way analogous to the $\mathcal{QL}$ soundness proof,
we prove the soundness of $\mathcal{CL}$ only by means of WDOL.

Recall Definitions \ref{exists-c} and \ref{one} for
``model,'' ``valid,'' and ``consequence.''

We now prove the soundness of classical logic by means of
WDOL, i.e., that if $A$ is a theorem in ${\mathcal{CL}}$,
then $A$ is valid in any WDOL model.

\begin{theorem}\label{th:soundness-c}{\rm [Soundness]}
$\qquad\Gamma\vdash A\quad\Rightarrow\quad\Gamma\vDash_\mathcal{M} A$
\end{theorem}

\begin{proof}
We must show that
any axiom A1--A4, given by Eqs.~(\ref{eq:cl-a1}--\ref{eq:cl-a4}),
is valid in any WDOL model $\mathcal{M}$, and that any set of formulas that
are consequences of $\Gamma$ in the model are closed under the rule of
inference
R1, Eq.~(\ref{eq:cl-r1}).

Let us put $a=h(A)$, $b=h(B)$, \ldots

By Theorem \ref{th:ba-sim}, we can prove that WDOL is
equal to OL restricted to all Boolean algebra
conditions of the form $t\equiv_0 s=1$, where
$t$ and $s$ are terms and
$t=s$ is an equation that holds in all Boolean algebras.
Notice that according to Corollary~\ref{cor:w-t-d},
$t\equiv_0 s=1$ is not generally equivalent to $t=s$ in WDOL.
For example, the mappings of
A1--A3 and R1 hold in every ortholattice, and the ortholattice
mapping of A4 does not make the ortholattice even orthomodular
let alone distributive. In other words,
\begin{eqnarray}
(a\cap(b\cup c))\equiv_0((a\cap b)\cup(a\cap c))=1
\end{eqnarray}
does not imply $(a\cap(b\cup c))=((a\cap b)\cup(a\cap c))$,
and therefore we cannot speak of distributivity within
$\mathcal{CL}$.\end{proof}

\section{The completeness of $\mathcal{QL}$ for WOML
models: non-orthomodularity confirmed}
\label{sec:compl-ql}

Our main task in proving the soundness of $\mathcal{QL}$ in the previous
section was to show that all axioms as well as the rules of inference
(and therefore all theorems) from $\mathcal{QL}$ hold in WOML. The task
of proving the completeness of $\mathcal{QL}$ is the opposite one:
we have to impose the structure of WOML on the set
${\mathcal{F}}^\circ$ of formulas of $\mathcal{QL}$.

We start with a relation of congruence, i.e.,
a relation of equivalence compatible with the operations in
$\mathcal{QL}$. We make use of an equivalence relation to
establish a correspondence between formulas of $\mathcal{QL}$ and
formulas of WOML. The resulting equivalence classes stand for elements
of a WOML and enable the completeness proof of $\mathcal{QL}$ by 
means of this WOML.

Our definition of congruence involves a special set of valuations 
on lattice O6 (shown in Figure \ref{fig:O6} in Section \ref{sec:latt}) 
called ${\mathcal{O}}${\rm 6} and defined as follows.  Its
definition is the same for both the quantum logic completeness proof in
this section and the classical logic completeness proof in Section
\ref{sec:compl-cl}.

\begin{definition}\label{D:hexagon}Letting {\rm O6} represent 
the lattice from Definition \ref{df:o6}, we define 
${\mathcal{O}}${\rm 6} as the set of all mappings 
$o:{\mathcal{F}}^\circ\longrightarrow {\rm O}6$ such that for
$A,B\in{\mathcal{F}}^\circ$,
$o(\neg A)=o(A)'$, and $o(A\vee B)=o(A)\cup o(B)$.
\end{definition}

The purpose of ${\mathcal{O}}${\rm 6} is to let us refine the
equivalence classes used for the completeness proof, so that the
Lindenbaum algebra will be a proper WOML, i.e. one that is not
orthomodular.  This is accomplished by conjoining the term $(\forall
o\in{\mathcal{O}}6)[(\forall X\in\Gamma)(o(X)=1) \Rightarrow o(A)=o(B)]$
to the equivalence relation definition, meaning that for equivalence we
require also that (whenever the valuations $o$ of the wffs in $\Gamma$ are
all 1) the valuations of wffs $A$ and $B$ map to the same point in the
lattice O6.  For example, the two wffs $A\vee B$ and $A\vee (\neg
A\wedge (A\vee B))$ will become members of two separate equivalence
classes by Theorem \ref{th:non-distr} below.  Without the conjoined
term, these two wffs would belong to the same equivalence class.  The
point of doing this is to provide a completeness proof that is not
dependent in any way on the orthomodular law, to show that completeness
does not require that the underlying models be OMLs.

\begin{theorem}\label{th:congruence-nonoml}
The relation of {\em equivalence} $\approx_{\Gamma,\mathcal{QL}}$
or just $\approx$, defined as
\begin{eqnarray}&&\hskip-5ptA\approx B\\&&\hskip8pt{\buildrel\rm
def\over =}\
\Gamma\vdash
A\equiv B\ \&\ (\forall o\in{\mathcal{O}}{\rm 6})[(\forall
X\in\Gamma)(o(X)=1)
\Rightarrow o(A)=o(B)],\nonumber
\label{eq:equiv-noml}
\end{eqnarray}
is a relation of congruence in the algebra
$\mathcal{F}$, where\/ $\Gamma\subseteq{\mathcal{F}}^\circ$
\end{theorem}

\begin{proof} Let us first prove that $\approx$ is an equivalence
relation. $\>A\approx A\>$ follows from A1
[Eq.~(\ref{eq:kalmb-A1})] of system $\mathcal{QL}$ and the identity
law of equality.
If $\Gamma\vdash A\equiv B$, we can detach the left-hand
side of A12 to conclude $\Gamma\vdash B\equiv A$, through the use of
A13 and repeated uses of A14 and R1.  From this and commutativity
of equality, we conclude $\>A\approx B\>\Rightarrow\>
B\approx A$.  (For brevity we will
not usually
mention further uses of A12, A13, A14, and R1 in what follows.)
 The proof of transitivity runs as
follows.
\begin{eqnarray}
A\approx B&&\quad\&\quad B\approx C\label{line1}\\
&&\Rightarrow\ \Gamma\vdash A\equiv   B\quad\&\quad \Gamma\vdash
B\equiv   C\nonumber\\
&&\hskip-20pt\&\ (\forall o\in{\mathcal{O}}6)
[(\forall X\in\Gamma)(o(X)=1)\ \Rightarrow\
o(A)=o(B)]\nonumber\\
&&\hskip-20pt\&\ (\forall o\in{\mathcal{O}}6)
[(\forall X\in\Gamma)(o(X)=1)\ \Rightarrow\
o(B)=o(C)]\nonumber\\
&&\Rightarrow\ \Gamma\vdash A\equiv   C \nonumber\\
&&\hskip-20pt\&\ (\forall o\in{\mathcal{O}}6)[(\forall
X\in\Gamma)(o(X)=1)\ \Rightarrow\
o(A)=o(B)\ \&\ o(B)=o(C)].\nonumber
\end{eqnarray}
In the last line above, we obtain $\Gamma\vdash A\equiv C$ 
(see Sec.~\ref{subsec:q-logic}) by using A2, A14 twice, and 
R1 six times and the last metaconjunction reduces to  
$\ o(A)=o(C)\ $ by transitivity of equality.
Hence the conclusion $A\approx C$ by definition.

In order to be a relation of congruence, the relation of
equivalence must be compatible with the operations $\neg$ and
$\vee$. These proofs run as follows.
\begin{eqnarray}
A\approx B&&\label{line-1}\\
&&\Rightarrow\Gamma\vdash A\equiv B\nonumber\\
&&\hskip-20pt\&\ \ (\forall o\in{\mathcal{O}}6)
[(\forall X\in\Gamma)(o(X)=1)\ \Rightarrow\ o(A)=o(B)]\nonumber\\
&&\Rightarrow\Gamma\vdash\neg A\equiv\neg B\nonumber\\
&&\hskip-20pt\&\ \ (\forall o\in{\mathcal{O}}6)
[(\forall X\in\Gamma)(o(X)=1)\ \Rightarrow\ o(A)'=o(B)']
\nonumber\\
&&\Rightarrow\Gamma\vdash\neg A\equiv\neg B\nonumber\\
&&\hskip-20pt\&\ \ (\forall o\in{\mathcal{O}}6)
[(\forall X\in\Gamma)(o(X)=1)\ \Rightarrow\ o(\neg A)=o(\neg B)]
\nonumber\\
&&\Rightarrow\neg A\approx\neg B\nonumber
\end{eqnarray}
\begin{eqnarray}
A\approx B&&\label{line-11}\\
&&\Rightarrow \Gamma\vdash A\equiv B\nonumber\\
&&\hskip-20pt\&\ \ (\forall o\in{\mathcal{O}}6)
[(\forall X\in\Gamma)(o(X)=1)\ \Rightarrow\ o(A)=o(B)]\nonumber\\
&&\Rightarrow\Gamma\vdash(A\vee C)\equiv(B\vee C)\nonumber\\
&&\hskip-30pt\&\ \ (\forall o\in{\mathcal{O}}6)
[(\forall X\in\Gamma)(o(X)=1)\ \Rightarrow\ o(A)\cup o(C)=o(B)\cup o(C)]
\nonumber\\
&&\Rightarrow(A\vee C)\approx(B\vee C)\nonumber
\end{eqnarray}
In the second step of Eq.~\ref{line-1}, we used A3.  In the
second step of Eq.~\ref{line-11}, we used A4 and A10.
For the quantified part of these expressions, we applied the definition
of ${\mathcal{O}}6$.
\end{proof}

\begin{definition}\label{D:equiv-class-sets-woml}
The equivalence class for wff $A$ under the relation of equivalence
$\approx$ is
defined as $|A|=\{B\in {\mathcal{F}}^\circ:A\approx B\}$, and we denote
${\mathcal{F}}^\circ/\!\approx\ =\{|A|:A\in {\mathcal{F}}^\circ\}$.
The equivalence classes define the natural morphism
$f:{\mathcal{F}}^\circ\longrightarrow
{\mathcal{F}}^\circ/\!\approx$, which gives
$f(A)\ =^{\rm def}\ |A|$. We write $a=f(A)$, $b=f(B)$, etc.
\end{definition}

\begin{lemma}\label{L:equality-non-q}
The relation $a=b$ on ${\mathcal{F}}^\circ/\!\approx$ is given by:
\begin{eqnarray}
\hskip80pt |A|=|B|\qquad&\Leftrightarrow&\qquad A\approx B
\label{eq:equation-non-om-q}
\end{eqnarray}
\end{lemma}

\begin{lemma}\label{L:lind-alg-non-q} The Lindenbaum algebra
${\mathcal{A}}=\langle {\mathcal{F}}^\circ/\!\approx,\neg/\!\approx,
\vee/\!\approx\rangle$ is a
{\rm WOML}, i.e., Eqs.~(\ref{eq:aub})--(\ref{eq:aAb})
 and Eq.~(\ref{eq:woml2}) hold for
$\neg/\!\approx$ and $\vee/\!\approx$
as  $'$ and $\cup$
respectively {\em [where---for simplicity---we use the same symbols
($'$ and $\cup$) as for O6, since there are no ambiguous
expressions in which the origin of the operations would not
be clear from the context]}.
\end{lemma}

\begin{proof}
For the $\Gamma\vdash A\equiv B$ part of the $A\approx B$ definition,
the proofs of the ortholattice conditions,
Eqs.~(\ref{eq:aub})--(\ref{eq:aAb}), follow from
A5, A6, A9, the dual of A8, the dual of A7, and DeMorgan's laws
respectively.
(The duals follow from DeMorgan's laws, derived from A10, A9, and A3.)
A11 gives us an analog of the OML law for the $\Gamma\vdash A\equiv B$
part, and the WOML law
Eq.~(\ref{eq:woml2}) follows from the OML law in an ortholattice.
For the quantified part of the $A\approx B$ definition,
lattice O6 is a WOML by Theorem \ref{th:O6}.
\end{proof}

\begin{lemma}\label{L:lind-alg-eq-1-q}
In the Lindenbaum algebra $\mathcal{A}$, if
$f(X)=1$ for all $X$ in\/ $\Gamma$ implies $f(A)=1$,
then\/ $\Gamma\vdash A$.
\end{lemma}

\begin{proof}
Let us assume that $f(X)=1$ for all $X$ in $\Gamma$ implies $f(A)=1$ i.e.
$|A|=1=|A|\cup |A|'=|A\vee\neg A|$, where the
first equality is from
Definition \ref{D:equiv-class-sets-woml}, the
second equality follows from Eq.~(\ref{D:onezero})
(the definition of 1 in an ortholattice), and
the third from the fact that $\approx$ is a congruence.
Thus
$A \approx (A\vee\neg A)$, which by definition means
$\Gamma\vdash
A\equiv (A\vee\neg A)\ \&\ (\forall o\in{\mathcal{O}}6)[(\forall
X\in\Gamma)(o(X)=1)
\Rightarrow o(A)=o((A\vee\neg A))]$. This implies, in particular,
$\Gamma\vdash
A\equiv (A\vee\neg A)$.  In any ortholattice, $a\equiv (a\cup a')=a$
holds.  By analogy, we can prove
$\Gamma\vdash (A\equiv (A\vee\neg A))\equiv A$ from $\mathcal{QL}$
axioms A1--A15.
Detaching the left-hand side (using A12, A13, A14, and R1), we
conclude $\Gamma\vdash A$.
\end{proof}

\begin{theorem}\label{th:non-distr}The orthomodular law does not hold
in $\mathcal{A}$.
\end{theorem}

\begin{proof} This is Theorem 3.27 from \cite{mpcommp99}, and 
the proof provided there runs as follows. 
We assume ${\mathcal F}^\circ$ contains at least two elementary
(primitive) propositions $p_0,p_1,\ldots$. We pick
a valuation $o$ that maps two of them, $A$ and $B$, to
distinct nodes $o(A)$ and $o(B)$ of O6 that are
neither 0 nor 1 such that $o(A)\le o(B)$
[i.e. $o(A)$ and $o(B)$ are on the same side of hexagon O6 in
Figure \ref{fig:O6} in Section \ref{sec:latt}]. 
From the structure of O6, we obtain
$\>o(A)\cup o(B)=o(B)$ and $o(A)\cup(o(A)'\cap(o(A)\cup o(B)))
=o(A)\cup(o(A)'\cap o(B))= o(A)\cup 0=o(A)$. Therefore
$o(A)\cup o(B)\ne o(A)\cup (o(A)' \cap (o(A)\cup o(B))$, i.e.,
$o(A\vee B)\ne o(A\vee (\neg A\wedge(A\vee B)))$.
This falsifies $(A\vee B)\approx (A\vee(\neg A\wedge (A\vee B))$.
Therefore $a\cup b\ne a\cup (a'\cap(a\cup b))$,
providing a counterexample to the orthomodular law 
for ${\mathcal F}^\circ/\!\approx$.
\end{proof}

\begin{lemma}\label{L:model-wdol}$\mathcal{M}=\langle
\mathcal{F}/\!\approx,f\rangle$ is a {\rm WOML} model.
\end{lemma}

\begin{proof} Follows from Lemma \ref{L:lind-alg-non-q}. \end{proof}

Now we are able to prove the completeness of $\mathcal{QL}$, i.e.,
that if a formula {\rm A} is a consequence of a set
of wffs $\Gamma$
in all {\rm WOML} models,
then $\Gamma\vdash A$.  In particular, when $\Gamma=\varnothing$,
all valid formulas are provable in $\mathcal{QL}$.  (Recall from
the note below Definition \ref{one} that
the left-hand side of the metaimplication below is implicitly
quantified over all WOML models $\mathcal{M}$.)

\begin{theorem}\label{th:completeness-woml}{\rm[Completeness]}
$\qquad\Gamma\vDash_\mathcal{M} A\qquad\Rightarrow\qquad
\Gamma\vdash A$.
\end{theorem}

\begin{proof}
$\Gamma\vDash_\mathcal{M} A$ means that in all WOML models
$\mathcal{M}$, if $f(X)=1$ for all $X$ in $\Gamma$,
then $f(A)=1$ holds.   In particular, it holds for
$\mathcal{M}=\langle
\mathcal{F}/\!\approx,f\rangle$, which is a WOML model
by Lemma \ref{L:model-wdol}.  Therefore, in the
Lindenbaum algebra $\mathcal{A}$, if
$f(X)=1$ for all $X$ in $\Gamma$, then $f(A)=1$ holds.
By Lemma \ref{L:lind-alg-eq-1-q}, it follows that
$\Gamma\vdash A$.
\end{proof}

\section{The completeness of $\mathcal{CL}$ for WDOL
models: non-distributivity confirmed}
\label{sec:compl-cl}

In this section we will prove the completeness of $\mathcal{CL}$,
i.e., we will impose the structure of WDOL on the set
${\mathcal{F}}^\circ$ of formulas of $\mathcal{CL}$.

We start with a relation of congruence, i.e.,
a relation of equivalence compatible with the operations in
$\mathcal{CL}$. We have to make use of an equivalence relation to
establish a correspondence between formulas from $\mathcal{CL}$ and
formulas from WDOL. The resulting equivalence classes stand for elements
of a WDOL and enable the completeness proof of $\mathcal{CL}$.

\begin{theorem}\label{th:congruence-nondist}
The relation of {\em equivalence} $\approx_{\Gamma,\mathcal{CL}}$
or just $\approx$, defined as
\begin{eqnarray}&&\hskip-5ptA\approx B\\&&\hskip8pt{\buildrel\rm
def\over =}\
\Gamma\vdash
A\equiv_0 B\ \&\ (\forall o\in{\mathcal{O}}6)[(\forall
X\in\Gamma)(o(X)=1)
\Rightarrow o(A)=o(B)],\nonumber
\label{eq:equiv-noml-c}
\end{eqnarray}
is a relation of congruence in the algebra
$\mathcal{F}$.
\end{theorem}

\begin{proof}
The axioms and rules of $\mathcal{QL}$, A1--A15 and R1,
i.e., Eqs.~(\ref{eq:kalmb-A1})--(\ref{eq:kalmb-R1}),
are theorems of $\mathcal{CL}$, A1--A4 and R1, i.e.
Eqs.~(\ref{eq:cl-a1})--(\ref{eq:cl-r1}). To verify this
we refer the reader to {\em Principia
Mathematica} by Alfred Whitehead and Bertrand Russell 
\cite{principia}, where the $\mathcal{QL}$ axioms either 
will be found as theorems or can easily be derived from them.  
For example, axiom A1 of $\mathcal{QL}$ is given as 
Theorem *4.2 \cite[p.~116]{principia} after using 
Theorem *5.23 \cite[p.~124]{principia} to convert from $\equiv_0$ to $\equiv$.
This will let us take advantage of parts of the completeness
proof for $\mathcal{QL}$, implicitly using Theorem *5.23 
\cite[p.~124]{principia} in either direction as required.

With this in mind, the proof that $\approx$ is an equivalence and
congruence relation  becomes exactly the proof of Theorem
\ref{th:congruence-nonoml}.
\end{proof}

\begin{definition}\label{D:equiv-class-sets-wdol}
The equivalence class for wff $A$ under the relation of equivalence
$\approx$ is
defined as $|A|=\{B\in {\mathcal{F}}^\circ:A\approx B\}$, and we denote
${\mathcal{F}}^\circ/\!\approx\ =\{|A|\in {\mathcal{F}}^\circ\}$.
The equivalence classes define the natural morphism
$f:{\mathcal{F}}^\circ\longrightarrow
{\mathcal{F}}^\circ/\!\approx$, which gives
$f(A)\ =^{\rm def}\ |A|$. We write $a=f(A)$, $b=f(B)$, etc.
\end{definition}

\begin{lemma}\label{L:equality-non-d}
 The relation $a=b$ on ${\mathcal{F}}^\circ/\!\approx$ is given as:
\begin{eqnarray}
\hskip80pt |A|=|B|\qquad&\Leftrightarrow&\qquad A\approx B
\label{eq:equation-non-om}
\end{eqnarray}
\end{lemma}

\begin{lemma}\label{L:lind-alg-non-d} The Lindenbaum algebra
${\mathcal{A}}=\langle {\mathcal{F}}^\circ/\!\approx,\neg/\!\approx,
\vee/\!\approx,\wedge/\!\approx\rangle$ is a
{\rm WDOL}, i.e., Eqs.~(\ref{eq:aub})--(\ref{eq:aAb})
 and Eq.~(\ref{eq:wdol2}), hold for
$\neg/\!\approx$ and $\vee/\!\approx$
as  $'$ and $\cup$
respectively.
\end{lemma}

\begin{proof}
For the $\Gamma\vdash A\equiv_0 B$ part of the $A\approx B$ definition,
the proofs of the ortholattice axioms are identical to those
in the proof of Lemma \ref{L:lind-alg-non-q} (after using
using Theorem *5.23 on p.~124 of Ref.~\cite{principia} to convert
between $\equiv_0$ and $\equiv$).
The WDOL law Eq.~(\ref{eq:wdol2}) for the $\Gamma\vdash A\equiv_0 B$
part can be derived using Theorems *5.24, *4.21, *5.17, *3.2, *2.11,
and *5.1 \cite[pp.~101--124]{principia}.
For the quantified part of the $A\approx B$ definition,
lattice O6 is a WDOL by Theorem \ref{th:O6}.
\end{proof}

\begin{lemma}\label{L:lind-alg-eq-1-c}
In the Lindenbaum algebra $\mathcal{A}$, if
$f(X)=1$ for all $X$ in\/ $\Gamma$ implies $f(A)=1$,
then\/ $\Gamma\vdash A$.
\end{lemma}

\begin{proof}
Identical to the proof of Lemma \ref{L:lind-alg-eq-1-q}.
\end{proof}

\begin{theorem}\label{th:non-distr-c}Distributivity does not hold
in $\mathcal{A}$.
\end{theorem}

\begin{proof}
$(a\cap(b\cup c))=((a\cap b)\cup(a\cap c))$ fails in O6. 
Cf.~the proof of Theorem \ref{th:non-distr}.
\end{proof}

\begin{lemma}\label{L:model-wdol-c}$\mathcal{M}=\langle
\mathcal{F}/\!\approx,f\rangle$ is a {\rm WDOL} model.
\end{lemma}

\begin{proof} Follows Lemma \ref{L:lind-alg-non-d}. \end{proof}

Now we are able to prove the completeness of $\mathcal{CL}$, i.e.,
that if a formula {\rm A} is a consequence of a set
of wffs $\Gamma$
in all {\rm WDOL} models,
then $\Gamma\vdash A$.  In particular, when $\Gamma=\varnothing$,
all valid formulas are provable in $\mathcal{QL}$.

\begin{theorem}\label{th:completeness-wdol-c}{\rm[Completeness]}
$\qquad\Gamma\vDash_\mathcal{M} A\quad \Rightarrow\quad \Gamma\vdash A$
\end{theorem}

\begin{proof}
Analogous to the proof of Theorem \ref{th:completeness-woml}.
\end{proof}

\section{The completeness of $\mathcal{QL}$ for OML models: 
orthomodularity regained}
\label{sec:compl-oml}

Completeness proofs for $\mathcal{QL}$ carried out in the 
literature so far---with the exception of Pavi\v ci\'c
and Megill \cite{mpcommp99}---do not invoke Definition 
\ref{df:o6} and Theorem \ref{th:O6}, and
instead of Theorem \ref{th:congruence-nonoml} one invokes the following one:

\begin{theorem}\label{th:congruence-q} Relation
$\approx$ defined as
\begin{eqnarray}
\hskip110pt A\approx B\ & {\buildrel\rm def\over =}\ &
\Gamma\vdash A\equiv B
\label{eq:equiv-c}
\end{eqnarray}
is a relation of congruence in the algebra $\mathcal{F}$.
\end{theorem}

Instead of Definition \ref{D:equiv-class-sets-woml} one has:

\begin{definition}\label{D:equiv-class}
The equivalence class under the relation of equivalence is
defined as $|A|=\{B\in {\mathcal{F}}^\circ:A\approx B\}$,
and we denote ${\mathcal{F}}^\circ/\!\approx\
= \{|A|\in {\mathcal{F}}^\circ\}$
The equivalence classes define the natural morphism
$f:{\mathcal{F}}^\circ\longrightarrow {\mathcal{F}}^\circ/\!\approx$,
which gives $f(A)\ =^{\rm def}\ |A|$. We write $a=f(A)$,
$b=f(A)$, etc.
\end{definition}

And instead of Lemma \ref{L:equality-non-q} one is able to obtain:

\begin{lemma}\label{L:equality-d}
The relation $a=b$ on ${\mathcal{F}}^\circ/\!\approx$ is given as:
\begin{eqnarray}
&&a=b\quad\Leftrightarrow\quad
|A|=|B|\quad\Leftrightarrow\quad A\approx B\quad\Leftrightarrow\quad
\Gamma\vdash A\equiv B
\label{eq:equation}
\end{eqnarray}
\end{lemma}

Hence, from the following easily provable theorem in $\mathcal{QL}$:
\begin{eqnarray}&&\vdash(A\equiv B)\equiv(C\vee\neg C)
\quad\Rightarrow\quad\vdash A\equiv B\label{eq:logical-distr}
\end{eqnarray}
one is also able to get:
\begin{eqnarray}&&
a\equiv b=1 \quad\Rightarrow\quad a=b\label{eq:hidden-oml}
\end{eqnarray}
in the Lindenbaum algebra $\mathcal{A}$, which is the
orthomodularity as given by Definition \ref{def:oml2}.~\cite{p98}

The point here is that Eq.~(\ref{eq:hidden-oml}) has nothing to do
with any axiom or rule of inference from $\mathcal{QL}$---it is nothing
but a consequence of the definition of the relation of equivalence from
Theorem \ref{th:congruence-q}. Hence, the very definition of the standard
relation of equivalence introduces a hidden axiom---the
orthomodu\-lar\-ity---into the Lindenbaum algebra  $\mathcal{A}$,
thus turning it into an orthomodular lattice. Without this hidden axiom,
the  Lindenbaum algebra stays WOML as required by the $\mathcal{QL}$
syntax. With it the Lindenbaum algebra turns into OML as follows.

\begin{lemma}\label{L:lind-alg-eq-1-q-2}
In the Lindenbaum algebra $\mathcal{A}$, if
$f(X)=1$ for all $X$ in\/ $\Gamma$ implies $f(A)=1$,
then\/ $\Gamma\vdash A$.
\end{lemma}

\begin{proof} In complete analogy to the proof of Theorem~\ref{L:lind-alg-eq-1-q}.
\end{proof}

\begin{theorem}\label{th:oml-2}The orthomodular law holds
in $\mathcal{A}$.
\end{theorem}

\begin{proof}
$a\cup(a'\cap(a\cup b))=a\cup b$ follows from A11, Eq.~(\ref{eq:a11})
and Eq.~(\ref{eq:hidden-oml}).
\end{proof}

\begin{lemma}\label{L:model-oml-2}$\mathcal{M}=\langle
\mathcal{F}/\!\approx,f\rangle$ is an {\rm OML} model.
\end{lemma}

\begin{proof} Follows from Lemma \ref{L:lind-alg-eq-1-q-2}. \end{proof}

Now we are able to prove the completeness of $\mathcal{QL}$, i.e.,
that if a formula {\rm A} is a consequence of a set of wffs $\Gamma$
in all {\rm OML} models, then $\Gamma\vdash A$.

\begin{theorem}\label{th:completeness-oml-2}{\rm[Completeness]}
$\qquad\Gamma\vDash_\mathcal{M} A\quad \Rightarrow\quad \Gamma\vdash A$
\end{theorem}

\begin{proof}
Analogous to the proof of Theorem \ref{th:completeness-woml}.
\end{proof}

\section{The completeness of $\mathcal{CL}$ for Boolean algebra models:
distributivity regained}
\label{sec:last}

The completeness proof carried out in almost all logic books and
textbooks do not invoke Definition \ref{df:o6},
Theorem \ref{th:O6}, and Theorem \ref{th:congruence-nondist}.
An exception is the {\em Classical and Nonclassical Logics} by Eric
Schechter \cite[p.~272]{schechter} who adopted them from Pavi\v ci\'c
and Megill \cite{mpcommp99} and presented in a reduced approach
which he called the {\em hexagon interpretation}.
Other books, though, are based on:

\begin{theorem}\label{th:congruence} Relation $\approx$ defined as
\begin{eqnarray}
\hskip110pt A\approx B\ & {\buildrel\rm def\over =}\ &
\Gamma\vdash A\equiv_0 B
\label{eq:equiv}
\end{eqnarray}
is a relation of congruence in the algebra $\mathcal{F}$.
\end{theorem}

Instead of Definition \ref{D:equiv-class-sets-wdol} one has:

\begin{definition}\label{D:equiv-class-c}
The equivalence class under the relation of equivalence is
defined as $|A|=\{B\in {\mathcal{F}}^\circ:A\approx B\}$,
and we denote ${\mathcal{F}}^\circ/\!\approx\
= \{|A|\in {\mathcal{F}}^\circ\}$
The equivalence classes define the natural morphism
$f:{\mathcal{F}}^\circ\longrightarrow {\mathcal{F}}^\circ/\!\approx$,
which gives $f(A)\ =^{\rm def}\ |A|$. We write $a=f(A)$,
$b=f(A)$, etc.
\end{definition}

And instead of Lemma \ref{L:equality-non-d} one is able to obtain:

\begin{lemma}\label{L:equality-d-c}
The relation $a=b$ on ${\mathcal{F}}^\circ/\!\approx$ is given as:
\begin{eqnarray}
&&a=b\quad\Leftrightarrow\quad
|A|=|B|\quad\Leftrightarrow\quad A\approx B\quad\Leftrightarrow\quad
\Gamma\vdash A\equiv_0 B
\label{eq:equation-c}
\end{eqnarray}
\end{lemma}

Hence, from the following easily provable theorem in $\mathcal{CL}$:
\begin{eqnarray}&&\vdash(A\equiv_0 B)\equiv_0(C\vee\neg C)
\quad\Rightarrow\quad\vdash A\equiv_0 B\label{eq:logical-distr-c}
\end{eqnarray}
one is also able to get:
\begin{eqnarray}&&
a\equiv_0 b=1 \quad\Rightarrow\quad a=b\label{eq:distributivity-c}
\end{eqnarray}
in the Lindenbaum algebra $\mathcal{A}$, which is the
distributivity as given by Definition \ref{def:ba2}. \cite{p98}
The point here is that Eq.~(\ref{eq:distributivity-c}) has nothing to do
with any axiom or rule of inference from $\mathcal{CL}$---it is nothing
but a consequence of the definition of the relation of equivalence from
Theorem \ref{th:congruence}. Hence, the very definition of the standard
relation of equivalence introduces the distributivity as a hidden axiom
into the Lindenbaum algebra $\mathcal{A}$ and turns it into a Boolean 
algebra.

\begin{theorem}\label{th:completeness-boole-2}{\rm[Completeness]}
$\qquad\Gamma\vDash_\mathcal{M} A\quad \Rightarrow\quad \Gamma\vdash A$
\end{theorem}

\begin{proof}
Analogous to the proof of Theorem \ref{th:completeness-wdol-c}.
\end{proof}

\section{Discussion}
\label{sec:discussion}

In the above sections, we reviewed the historical results
that we considered relevant to decide whether quantum logic
can be considered a logic or not. In the Introduction, we
showed that many authors in the past thirty years tried to
decide on this question by starting with particular
models and their syntax---the orthomodular lattice for
quantum logic and Boolean algebra for classical. They
compared the models and often came to a conclusion that
since they are so different, quantum logic should not be
considered a logic. This was, however, in obvious
conflict with the growing number of well-formulated
quantum logic systems over the same period. We mentioned
some of them in the Introduction.

Orthomodular lattices and Boolean algebras {\em are} very different.
As reviewed in the Introduction, in any orthomodular lattice
all operations, variables, and constants are sixfold defined
(five {\em quantum} and one {\em classical}), and in a Boolean
algebra they all merge to classical operations, variables,
and constants (0,1). Both an orthomodular lattice and a Boolean
algebra can be formulated as equational systems---as reviewed in
Section \ref{sec:latt}. Such equational systems can mimic both 
quantum and classical logics and show that one can formulate 
the Deduction Theorem in a special orthomodular
lattice---a distributive one, i.e., a Boolean algebra---but
cannot in a general one. As a consequence, the operation
of implication---which the Deduction Theorem\footnote{See
footnote No.~{\protect{\ref{foot:d-t}}}.} is based
on---plays a special unique role in classical logic and
does not in quantum logic. Also, the Boolean algebra used
as a model for classical logic is almost always two-valued,
i.e., it consists of only two elements 0 and 1, and an
orthomodular lattice, according to the Kochen-Specker theorem,
cannot be given a $\{0,1\}$ valuation.\footnote{In 2004 Mladen
Pavi{\v c}i{\'c}, Jean-Pierre Merlet, Brendan Mc{K}ay,
and Norman Megill gave exhaustive algorithms for
generation of Kochen-Specker vector systems
with arbitrary number of vectors in Hilbert spaces of
arbitrary dimension.~\cite{pmmm04b,pmmm03a,pavicic-book-05}
The algorithms use MMP (McKay-Megill-Pavi\v ci\'c) diagrams
for which in 3-dim Hilbert space a direct correspondence to
Greechie and Hasse diagrams can be established. Thus, we
also have a constructive proof within the lattice itself.}

So, recently research was carried out on whether a logic
could have more than one model of the same type, e.g., an
ortholattice, with the idea of freeing logics of any
semantics and valuation. The result was affirmative, and
a consequence was that quantum logic can be considered
a logic in the same sense in which classical logic
can be considered a logic. The details are given in
Sections~{$\ref{sec:logic}\mbox{--}\ref{sec:last}$}, where we chose
Kalmbach's system to represent quantum logic in
Section \ref{subsec:q-logic} and Hilbert and Ackermann's
presentation of {\em Principa Mathematica} to represent 
classical logic in Section \ref{subsec:cl-logic} (although we
could have chosen any other system mentioned in the
Introduction or from the literature).\footnote{Quantum logics 
given by Mladen Pavi\v ci\'c \cite{pav89} and 
by Mladen Pavi\v ci\'c and Norman Megill \cite{mpcommp99}
are particulary instructive since they contain only axioms  
designed so as to directly map into WOML conditions.} 

In Sections \ref{sec:sound-q} and \ref{sec:compl-ql}, we then
proved the soundness and completeness, respectively, of
quantum logic $\mathcal{QL}$ for a non-orthomodular model
WOML and in Sections \ref{sec:sound-c} and \ref{sec:compl-cl}
the soundness and completeness, respectively, of classical
logic $\mathcal{CL}$ for a non-distributive model WDOL.
Hence, with respect to these models,  quantum logic
$\mathcal{QL}$ cannot be called orthomodular and
classical logic $\mathcal{CL}$ cannot be called distributive
or Boolean. Also, neither $\mathcal{QL}$ nor $\mathcal{CL}$
can have a numerical valuation in general, since
the truth table method is inapplicable within
their OML, WOML, and WDOL models.

One might be tempted to ``explain'' these results in the
following way. ``It is true that WOML and WDOL obviously
contain lattices that violate the orthomodularity law,
for example the O6 hexagon (shown in Figure \ref{fig:O6}
in Section \ref{sec:latt}) itself, but most probably they
also {\em must} contain lattices that pass the law and
that would, with reference to Theorem \ref{th:oml-sim},
explain why we were able to prove the completeness of
quantum and classical logic for WOML and WDOL.'' This is,
however, not the case. We can prove the soundness and
completeness of quantum and classical logics using a class
of WOML lattices none of which pass the orthomodularity
law.~\cite{pmjlc08} Moreover, Eric Schechter has simplified
the results of Pavi\v ci\'c and Megill \cite{mpcommp99}
to the point of proving the soundness and completeness
of classical logic for nothing but O6 
itself.~\cite[p.~272]{schechter}

One of the conclusions Eric Schechter has drawn from the
unexpected non-distributivity of the WDOL models, especially
when reduced to the O6 lattice alone, is that all the axioms
that one can prove by means of $\{0,1\}$ truth tables, one can also
prove by any Boolean algebra, and by O6. So, logics are, first
of all, axiomatic deductive systems. Semantics are a next layer
that concern models and valuations. Quantum and classical logics
can be considered to be two such deductive systems. There are no
grounds for considering any of the two logics more ``proper''
than the other. As we have shown above, semantics of the logics
that consider their models show bigger differences
between the two aforementioned classical models than
between two corresponding quantum and classical models.

Whether we will ever use O6 semantics of classical logic or
WOML semantics of quantum logic remains an open question, but
these semantics certainly enrich our understanding of the role
of logics in applications to mathematics and physics. We cannot
make use of bare axiomatics of logic without specifying semantics
(models and valuations) for the purpose. By making such a choice
we commit ourselves to a particular model and disregard the
original logical axioms and
their syntax. Thus we do not use quantum logic itself in quantum
mechanics and in quantum computers but instead an orthomodular
lattice, and we do not use classical logic in our computers today but
instead a two-valued Boolean algebra (we even hardly ever use more
complicated Boolean algebras). We certainly cannot use O6 semantics
to build a computer or an arithmetic; however, one day we might
come forward with significant applications of these
alternative semantics, and then it might prove important to
have a common formal denominator for all the models---logics
they are semantics of. We can also impement an alternative
scenario---searching for different ortholattice semantics of the 
same logics.~\cite{pmjlc08}

Whatever strategy we choose to apply, we should always 
bear in mind that the syntaxes of the logics correspond to 
WOML, WDOL, and O6 semantics (models) while OML and Boolean 
algebra semantics (models) are imposed on the logics with the 
help of ``hidden'' axioms, Eqs.~(\ref{eq:hidden-oml}) and 
(\ref{eq:distributivity-c}), that emerge from the standard 
way of defining the relation of equivalence in the completeness 
proofs, Theorems \ref{th:congruence-q} and \ref{th:congruence}, 
of the logics for the latter models.

\bigskip

{\parindent=0pt
{\bf Acknowledgements}

\medskip
Supported by the Ministry of
Science, Education, and Sport of Croatia through the project
No.\ 082-0982562-3160.}


\bibliographystyle{plain}

\end{document}